\documentclass[aps,pra,twocolumn,superscriptaddress,showpacs,showkeys,superscriptaddress]{revtex4-2}
\usepackage{dcolumn}    
\usepackage{ifpdf}
\usepackage{amssymb,lineno,amsfonts}
\usepackage{graphicx}   
\usepackage{dcolumn}    
\usepackage{bm}         
\usepackage{mathrsfs}
\usepackage{upgreek}
\usepackage{mathtools}
\usepackage{epstopdf}
\usepackage{setspace}
\usepackage{hyperref}
\usepackage{float}
\usepackage{amsmath}
\usepackage{placeins}
\usepackage{bbm}
\usepackage[normalem]{ulem}
\usepackage[utf8]{inputenc}
\usepackage[usenames,dvipsnames]{xcolor}
\usepackage[mathscr]{euscript}
\usepackage[]{qcircuit}
\usepackage[normalem]{ulem}
\usepackage{physics}

\definecolor{med-blue}{RGB}{25,25,112}
\hypersetup{colorlinks, linkcolor={red},citecolor={blue}, urlcolor={MidnightBlue}}

\begin{document}
	\title{Physics-informed neural network for quantum control of NMR registers}

\author{Priya Batra}
\email{priya.batra@students.iiserpune.ac.in}

\author{T. S. Mahesh }
\email{mahesh.ts@iiserpune.ac.in}
\affiliation{Physics Department and NMR Research Center, 
		Indian Institute of Science Education and Research, Pune 411008, India}
		
\begin{abstract}
{Classical and quantum machine learning are being increasingly applied to various tasks in quantum information technologies. Here, we present an experimental demonstration of quantum control using a physics-informed neural network (PINN).  PINN's salient feature is how it encodes the entire control sequence in terms of its network parameters. This feature enables the control sequence to be later adopted to any hardware with optimal time discretization, which contrasts with conventional methods involving a priory time discretization.  Here, we discuss two important quantum information tasks: gate synthesis and state preparation. First, we demonstrate quantum gate synthesis by designing a two-qubit CNOT gate and experimentally implementing it on a heteronuclear two-spin NMR register.  Second, we demonstrate quantum state preparation by designing a control sequence to efficiently transfer the thermal state into the long-lived singlet state and experimentally implement it on a homonuclear two-spin NMR register. We present a detailed numerical analysis of the PINN control sequences regarding bandwidth, discretization levels, control field errors, and external noise.
}
\end{abstract}
		

\maketitle

\section{Introduction}
 \label{Introduction}
Efficient and robust quantum control is one of the critical needs for the practical realization of quantum technologies \cite{Acín_2018,dong2023learning}. It deals with the problem of efficient state-independent unitary synthesis or optimal transfer of one quantum state to another. Given its importance in a wide variety of problems, the development of quantum control methodologies is an active area of research, and numerous optimization algorithms have been designed and demonstrated \cite{khaneja2005, nielsen2007optimal, caneva2011a, koch2022, mahesh2023}. 

The use of machine learning protocols for different quantum problems has been a point of significant attention recently \cite{mehta_high-bias_2019,krenn_artificial_2023,Batra2021efficient}. Physics-informed neural networks (PINNs) are machine-learning algorithms that
involve built-in equations pertaining to certain physical laws \cite{raissi_physics_2017}.  After successfully implementing a neural network for solving Navier Stokes's equation \cite{raissi2018,raissi2020}, PINN started to gain attention for solving various other types of differential equations \cite{raissi_physics_2017,raissi_physics-informed_2019}. It is said to be a universal function approximator \cite{hornik1989}. Recently, it has been proposed to solve Schr\"odinger's equation \cite{gungordu2022a} as well as the open quantum system Lindbladian equation \cite{PhysRevLett.132.010801}. 

An important strength of the PINN algorithm is that it utilizes smooth basis functions to convert network parameters into the control function. Therefore, it generates smooth controls with low bandwidth that are easier to implement in experimental hardware \cite{gungordu2022a}.  Hence, PINN avoids the need for any pre-discretization prevalent in other quantum control algorithms \cite{khaneja2005, 10.1063/1.2903458, PhysRevResearch.2.013314}.  

This work demonstrates PINN-based quantum control for both quantum gate synthesis and quantum state preparation.  
First, we utilize PINN to design a two-qubit CNOT gate and experimentally implement it on a heteronuclear two-qubit NMR register.  We then utilize PINN to design a control sequence to efficiently transfer the thermal magnetization into the long-lived singlet state (LLS) and experimentally implement the sequence on a homonuclear two-qubit NMR register.  In both cases, we make detailed numerical analyses of the control sequences in terms of bandwidth, discretization levels, external noise, and control errors.

In the next section, we explain the theoretical ideas behind PINN. In Sec. \ref{sec:gs} and \ref{sec:s2s}, we describe the PINN-based CNOT gate and PINN-based LLS preparation, respectively.  Finally, we conclude in Sec. \ref{sec:summary}.

\section{Physics-informed neural network (PINN)}
Consider the total Hamiltonian $H(t)$ consisting of two components, a constant part $H_0$ and a time-dependent part such that 
\begin{align}
	H(t) = H_0 +  \sum_{k=0}^{M} u_{k,x}(t) I_{k,x} + u_{k,y}(t) I_{k,y},
	\label{eq:ham}
\end{align}
where $u_{k,x(y)}(t)$ are the $x(y)$ time-dependent control amplitudes on the $k$th nuclear species and $I_{k,x(y)} = \sum_m I_{k,x(y)}^{(m)}$ are the collective spin operators summed over spins of the same isotope. The task here is to find $u_{k,x(y)}(t)$ to achieve a target unitary $U_t$ or a target state $\rho_t$ efficiently. For this purpose, we utilize a deep learning-based physics-informed neural network (PINN) \cite{raissi_physics_2017,raissi_physics-informed_2019}. We implemented our neural network with JAX's Haiku library \cite{jax2018github,haiku2020github}.

\begin{figure}
	\begin{center}
		\includegraphics[trim=1cm 1cm 3.75cm 1.5cm, clip=true, width=9cm]{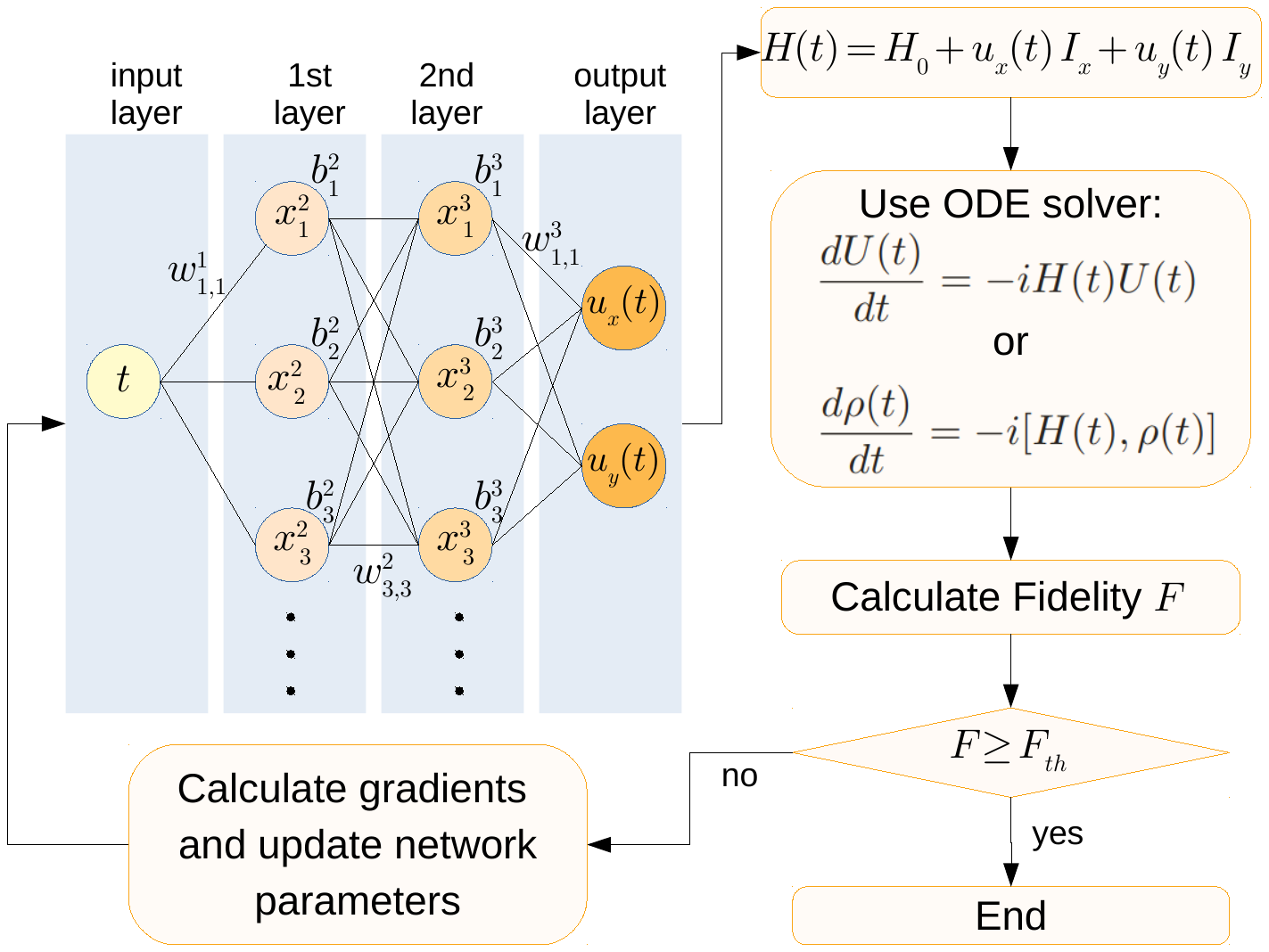} 
		\caption{Deep neural network and the flowchart describing different steps of PINN.  Here $x_i^l$ and $b_i^l$ denote the network nodes and their biases for the of $l$th layer, $w^l_{i,j}$ are the weights connecting $i$th node of $l$th layer with $j$th node of $(l+1)$th layer.  The first layer (input layer) has a single node which is the time parameter $t$ itself, i.e., $x^1_1 = t$.  The last layer (output layer) encodes the time-dependent control parameters $u_x(t) = x^{4}_1$ and $u_y(t) = x^4_2$ for a single isotope.  The 2nd and 3rd layers are the hidden layers of the (1,3,3,2) neural network explicitly shown here.   }
		\label{fig:NNandproc}
	\end{center}
\end{figure}

We employ a feed-forward network, as shown in Fig. \ref{fig:NNandproc}, with the time $t$ as input and control amplitudes $u_{k,x}(t)$ as output of the network. 
The node values are calculated according to
\begin{align}
x^{l+1}_{j} = \Gamma \left( \sum_i w^l_{i,j} \cdot x^{l}_i + b^{l+1}_j \right),
\end{align}
where the activation function $\Gamma$ is a nonlinear function that helps efficiently explore the parameter space. We have used
$\Gamma(\alpha) = \tanh{\alpha}$ as the activation function. 

It is important to notice that the control parameters are functions of time and network parameters. We need to iteratively optimize the network to get the best controls for any desired target unitary $U_t$ or state $\rho_t$. In every iteration, we inject the Hamiltonian $H(t)$ into Schr\"odinger equation (in $\hbar=1$ unit)
\begin{align}
\frac{dU(t)}{dt} = -iH(t)U(t),
\end{align}
and solve for the overall unitary $U(T)$ using JAX's ordinary differential equation solver \cite{deepmind2020jax}. Here we use $U(0) = \mathbbm{1}.$
Starting with a set of random network parameters, we calculate the gate-fidelity  
\begin{align}
	F_U &= \abs{\Tr(U_t^{\dagger} U(T))}^2.
\label{eq:gtfid}
\end{align}
In the case of state-to-state transfer, we use the von Neumann equation
\begin{align}
\frac{d\rho(t)}{dt} = -i[H(t),\rho(t)]
\label{eq:vne}
\end{align}
and solve for the final achieved state $\rho(T)$ by inputting a definite initial state $\rho(0) = \rho_i$.  Here, we use the state-fidelity
\begin{align}
F_\rho & = \Tr(\rho_t\rho(T)).
	\label{eq:stfid}
\end{align}
In either case, the goal is to maximize fidelity by optimizing the network. 
If the fidelity is less than a threshold value $F_{th}$, then we calculate gradients 
$\nabla_{w}F$ and $\nabla_{b}F$ for each weight and bias parameter using the auto differentiation (autodiff) algorithm in the Optax module \cite{10.5555/3122009.3242010}.
The updated parameters for the next iteration are calculated by adding the gradients scaled by a step parameter.  We optimize the step parameter using the ADAM optimizer.
The entire process is iterated, as shown in Fig. \ref{fig:NNandproc} until the fidelity is maximized.


\section{Gate synthesis: generating CNOT}
\label{sec:gs}
\begin{figure}
	\centering
	\includegraphics[trim=0cm 5.0cm 1cm 3cm,width=9cm,clip=]{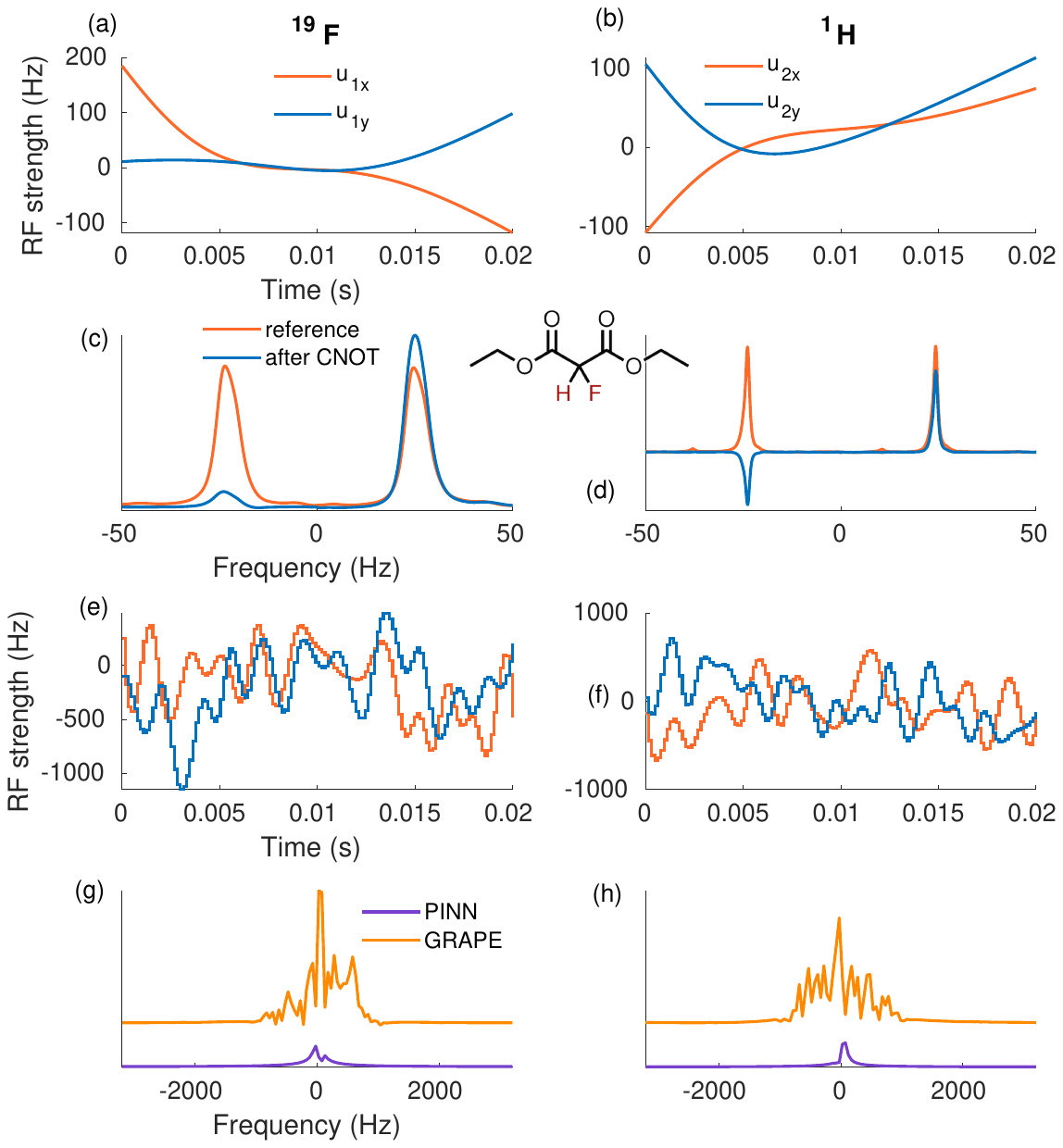}
	\caption{The $x$ and $y$ amplitudes of the PINN-CNOT on ${}^{19}$F (a) and ${}^{1}$H (b).  The reference NMR spectra and the spectra after CNOT for  ${}^{19}$F (c) and ${}^{1}$H (d).  The $x$ and $y$ amplitudes of GRAPE-CNOT on ${}^{19}$F (e) and ${}^{1}$H (f). The Fourier transforms of the PINN and GRAPE sequences on ${}^{19}$F (g) and ${}^{1}$H (h) showing relatively lower bandwidth of the PINN sequence. }
	\label{PINN_CNOT}
\end{figure}

Single-qubit gates and the two-qubit CNOT gate form a universal set of gates for quantum computation \cite{Nielsen_Chuang_2010}.  As a proof-of-principle demonstration, we employ PINN to prepare the CNOT gate and implement it on the two-qubit NMR register diethylfluoromalonate (DEFM) (shown in the inset of Fig. \ref{PINN_CNOT}) dissolved in solvent CDCl$_3$.  Here ${}^1$H and ${}^{19}$F nuclei constitute the two qubits.  All experiments are carried out on a high-resolution 500 MHz NMR spectrometer at an ambient temperature of 300 K.  In on-resonant doubly rotating frame precessing at the individual Larmor frequencies, the internal Hamiltonian for the system is 
\begin{align}
H_0 = 2 \pi J I_{1z} I_{2z},
\end{align}
where $J = 48.2$ Hz is the scalar coupling and $I_{1z}$, $I_{2z}$ are respectively the $z$-components of F and H spin operators. The control RF Hamiltonian is of the following form
\begin{align}
H_\mathrm{RF}(t) = u_{1x}(t) I_{1x} + u_{2x}(t) I_{2x} + u_{1y}(t) I_{1y} + u_{2y}(t) I_{2y},
\end{align}
with $u_{1(2)x(y)}$ being the $x (y) $ control amplitudes for qubits $1(2)$.  The total Hamiltonian is thus
\begin{align}
    H(t) = H_0 + H_\mathrm{RF}(t).
\end{align}

The PINN-generated RF controls for CNOT with $^{19}$F as the control qubit and $^1$H as the target qubit are shown in Fig. \ref{PINN_CNOT} (a,b).  
Here, we used (1,40,40,4) PINN with total time set to $T=20$ ms.
The experimental $^{19}$F and $^1$H spectra after the CNOT gate, along with the reference spectra (with $\pi/2$ detection pulse), are shown in Fig. \ref{PINN_CNOT} (c,d).

\begin{figure}
	\centering
	\includegraphics[trim=6cm 6cm 6cm 6cm,width=9cm,clip=]{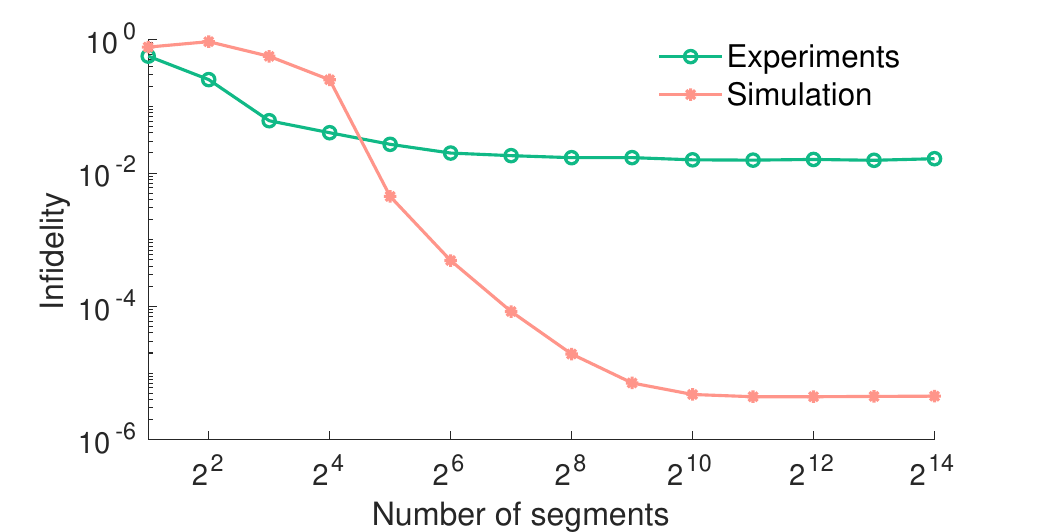}
	\caption{The infidelity ($1-F_U$) of the PINN-CNOT sequence with varying discretization size. The experimental infidelity hardly improves for beyond $2^7$ segments, while the simulated infidelity saturates after $2^10$ segments.}
	\label{PINN_CNOTdiscr}
\end{figure}

One major benefit of PINN is that the shape of the control profile is encoded as a nonlinear function of the network parameters, unlike discrete jumps in controls obtained by other algorithms such as GRAPE.  This has two advantages. First, the absence of rapid oscillations means smooth control profiles with minimum bandwidth.  To illustrate this, we compare the bandwidth with a GRAPE control sequence. Fig. \ref{PINN_CNOT} (e,f) shows a GRAPE-CNOT sequence (with $2^7$ segments), and Fig. \ref{PINN_CNOT} (g,h) compares the Fourier transforms of PINN and GRAPE sequences.  It is clear that the PINN sequence has a much lower bandwidth compared to the GRAPE sequence.
The second advantage is that PINN does not require a priory time-discretization.  Once optimized, the neural network can output the control sequence with any time discretization. This is highly useful for generating common control profiles and adopting them appropriately for multiple platforms with different time resolutions.  To illustrate this fact, we now discretize the PINN-CNOT sequence with varying segments and analyze its infidelity numerically and experimentally.  The results shown in Fig. \ref{PINN_CNOTdiscr} confirm that very low discretization, with smaller than $2^5$ segments, can not faithfully capture the control profiles, thereby leading to higher infidelities.  At $2^7$ segments, the experimental infidelity is close to $10^{-2}$, indicating a good performance of the CNOT gate. However, for discretizations better than $2^7$ segments, the experimental infidelities show no significant improvement.  In other control methods, such as GRAPE, one needs to run the whole optimization again to improve discretization.  The PINN algorithm avoids under or over-discretizations by capturing the control profiles in the network parameters, which can output any suitable discretization depending on the hardware requirement.

\section{State-to-state transfer: preparation of LLS}
\label{sec:s2s}
Long-lived singlet (LLS) order corresponds to the population difference between  the antisymmetric singlet state $\ket{S_0} = \frac{\ket{01} - \ket{10}}{\sqrt{2}}$ and the symmetric triplet state
$\ket{T_0} = \frac{\ket{01} + \ket{10}}{\sqrt{2}}$ in a pair of spin-1/2 nuclei.  For the isotropic Hamiltonian $-\mathbf{I}_1\cdot \mathbf{I}_2$, the state $\ket{T_0}$ forms a triply-degenerate eigenspace together with other symmetric triplet states $\ket{T_+} = \ket{00}$ and $\ket{T_-} = \ket{11}$, while $\ket{S_0}$ being the nondegenerate ground eigenstate. In the absence of symmetry breaking interactions, such as the chemical shift interaction, it is established that LLS can far outlive the single-spin relaxation time-constants $T_1$ and $T_2$ \cite{PhysRevLett.92.153003}. The survival of quantum coherence in LLS for such long durations has opened a wide range of applications from high-precision spectroscopy \cite{10.1039/9781788019972}, quantum computing \cite{PhysRevA.82.052302}, magnetic resonance imaging \cite{doi:10.1073/pnas.0908123106}, etc.  Numerous methods have been developed to transfer thermal state to LLS efficiently \cite{doi:10.1021/ja0490931,feng2012accessing,PhysRevLett.111.173002,PRAVDIVTSEV201656,KHURANA20178,PhysRevResearch.2.013314,10.1063/5.0159448}.  Here we demonstrate the use of PINN to prepare LLS starting from thermal magnetization efficiently.

\subsection{Experimental preparation of LLS by PINN}
\begin{figure}
	\centering
	\includegraphics[trim=1.4cm 2cm 0.5cm 1.5cm,width=9cm,clip=]{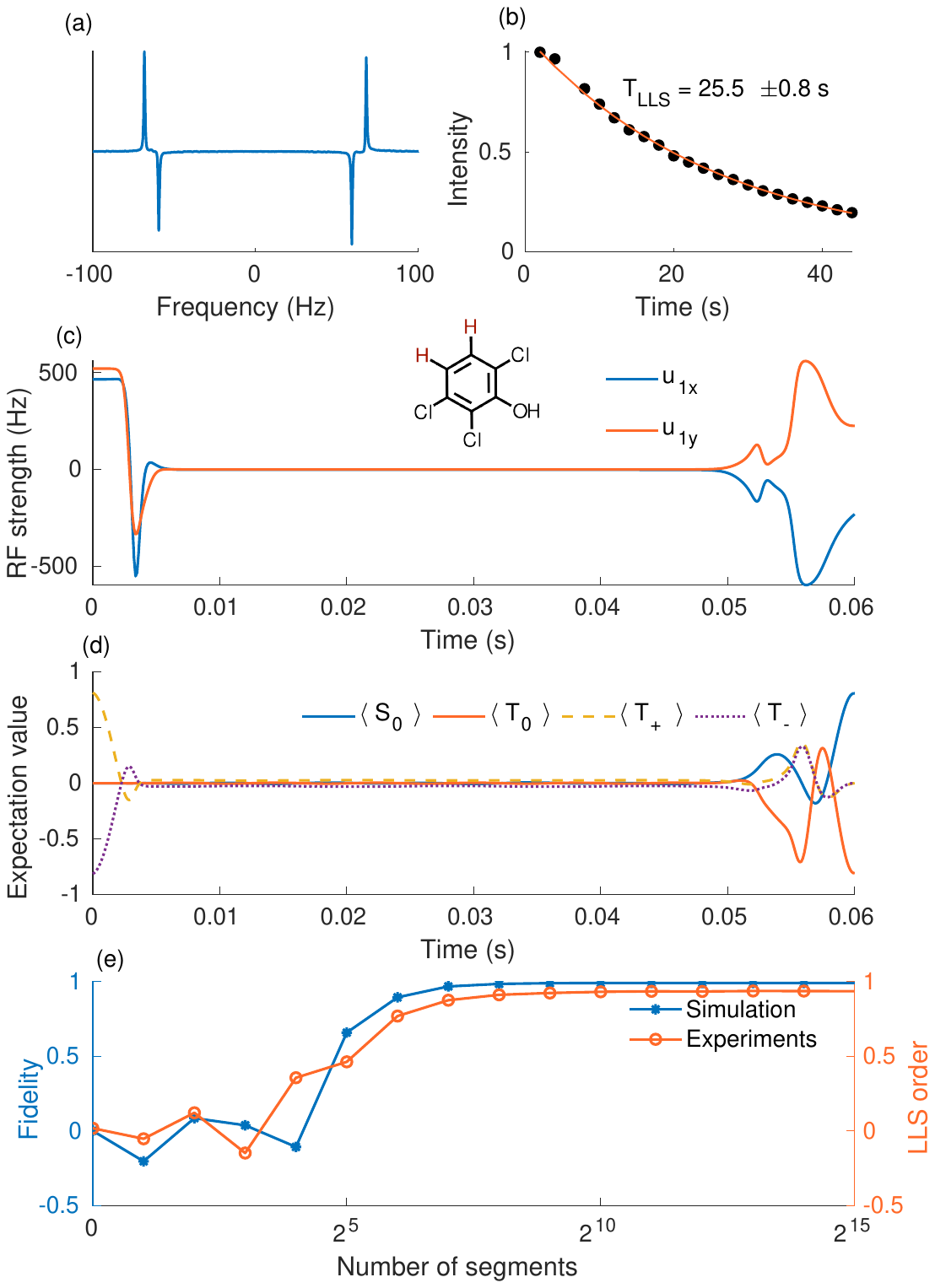}
	\caption{(a) LLS spectrum (b) Pulse sequence, $x$ and $y$ control amplitudes for ${}^1 H$ as a function of time (c) Decay of singlet magnetization under WALTZ 16 spin-lock.  The T$_1$ values of the two protons are 5.5 s and 5.6 s.  Thus, LLS outlives the single-spin non-equilibrium magnetizations by a factor of around 5. (d) The progression of the expectation value of all four eigenstates of the singlet-triplet basis. (e) Effect of discretization on continuous time pulse in simulation as well as experiments.}
	\label{PINN_LLS}
\end{figure}

We prepare LLS on the two protons of the 2-qubit sample 2,3,6-trichlorophenol (TCP) (shown in the inset of Fig.  \ref{PINN_LLS} (c)) dissolved in (CD$_3$)$_2$SO. The internal Hamiltonian of the two-qubit system is given by
\begin{align}
	H_0 = 2 \pi J I_{1z} I_{2z} -  \pi \Delta I_{1z} +  \pi \Delta I_{2z},
\end{align}
where coupling between two qubits is $J = 8.75$ Hz and chemical shift difference is $\Delta  =  127.5$ Hz. The external RF field has the following Hamiltonian
\begin{align}
	H_{RF}(t) = u_x(t) (I_{1x} + I_{2x}) + u_y(t) (I_{1y} + I_{2y}).
\end{align}
Thus, the total Hamiltonian is
\begin{align}
    H(t) = H_0 + H_{RF}(t).
\end{align}

Here, we used (1,60,60,60,2) PINN with three hidden layers.  The PINN sequence thus generated is shown in Fig. \ref{PINN_LLS} (c). Since the LLS order is not directly observable, we insert a delay $1/(4\Delta)$ followed by a $\pi/2$-pulse to convert it to the observable single-quantum magnetization. The resulting characteristic anti-phase LLS spectrum is shown \ref{PINN_LLS} (a).
To measure the lifetime of LLS, we stored it under the isotropic Hamiltonian realized by the WALTZ-16 spin-lock for up to $44$ sec.  The decay during the LLS storage plotted in Fig \ref{PINN_LLS} (b) yields the time-constant $T_\mathrm{LLS} = 25.5$ which is about five times the single-spin $T_1$ values $5.5$ s and  $5.6$ s for the two protons.  

To understand the workings of the PINN sequence, we analyze the evolution of all four eigenstates of the singlet-triplet basis.  The progressions of expectation values of these states during the PINN sequence are plotted in Fig. \ref{PINN_LLS} (d).  The initial thermal state has descending populations from $\ket{00}=\ket{T_+}$ to $\{\ket{01},\ket{10}\}=\frac{1}{\sqrt{2}}(\ket{T_0}\pm\ket{S_0})$ to $\ket{11} = \ket{T_-}$.  Interestingly, after an initial oscillation, the PINN sequence keeps all four expectation values close to zero until the last 10 ms, during which oscillations pick up again and population imbalance between $\ket{S_0}$ and  $\ket{T_0}$, corresponding to high LLS content, is achieved.

Like in the previous case, here also we analyze the effect of discretization level on the performance of the PINN-LLS sequence.  After the neural network is optimized, we can discretize the control sequence to any number of segments, which we have varied from $2^0$ to $2^{15}$, and analyze their performance both by simulations and experiments.  The results are shown in Fig. \ref{PINN_LLS} (e).  While the performance was seriously reduced for up to $2^5$ due to under-discretization, the sequence gave good fidelity, and a high LLS order was prepared at $2^7$ segments, which hardly improved thereafter.

\subsection{Robustness against noise}
It is important for any quantum control method to consider practical limitations such as external noise.  For example, one of the potential applications of LLS is in magnetic resonance imaging (MRI), where the longevity of LLS and hyperpolarization help improve the resolution, contrast, and sensitivity \cite{https://doi.org/10.1002/mrm.24430}.  In such applications, the nuclear spin pair in an endogenous molecule experiences high external noise, which the state-preparation sequence should account for.
To this end, instead of Eq. \ref{eq:vne}, we solve the Lindbladian equation,
\begin{align}
\frac{d \rho(t)}{dt} &= -i \, [H(t), \rho(t)] 
\nonumber \\ 
&~~+ \gamma \norm{H_0} \sum_{n}  \biggl(V_n \, \rho(t) \, V_n^\dagger - \frac{1}{2} \{\rho(t),V_n^\dagger V_n\}\biggr),
\end{align}
where $\rho(t) = U(t)\rho(0)U^\dagger(t)$,  $V_n$ are the noise operators, and
we have assumed uniform noise coefficient  $\gamma \norm{H_0}$ scaled by the norm $\norm{H_0}$ of $H_0$.

\begin{figure}[h!]
	\centering
	\includegraphics[trim=1cm 0cm 1cm 1.5cm,width=8.5cm,clip=]{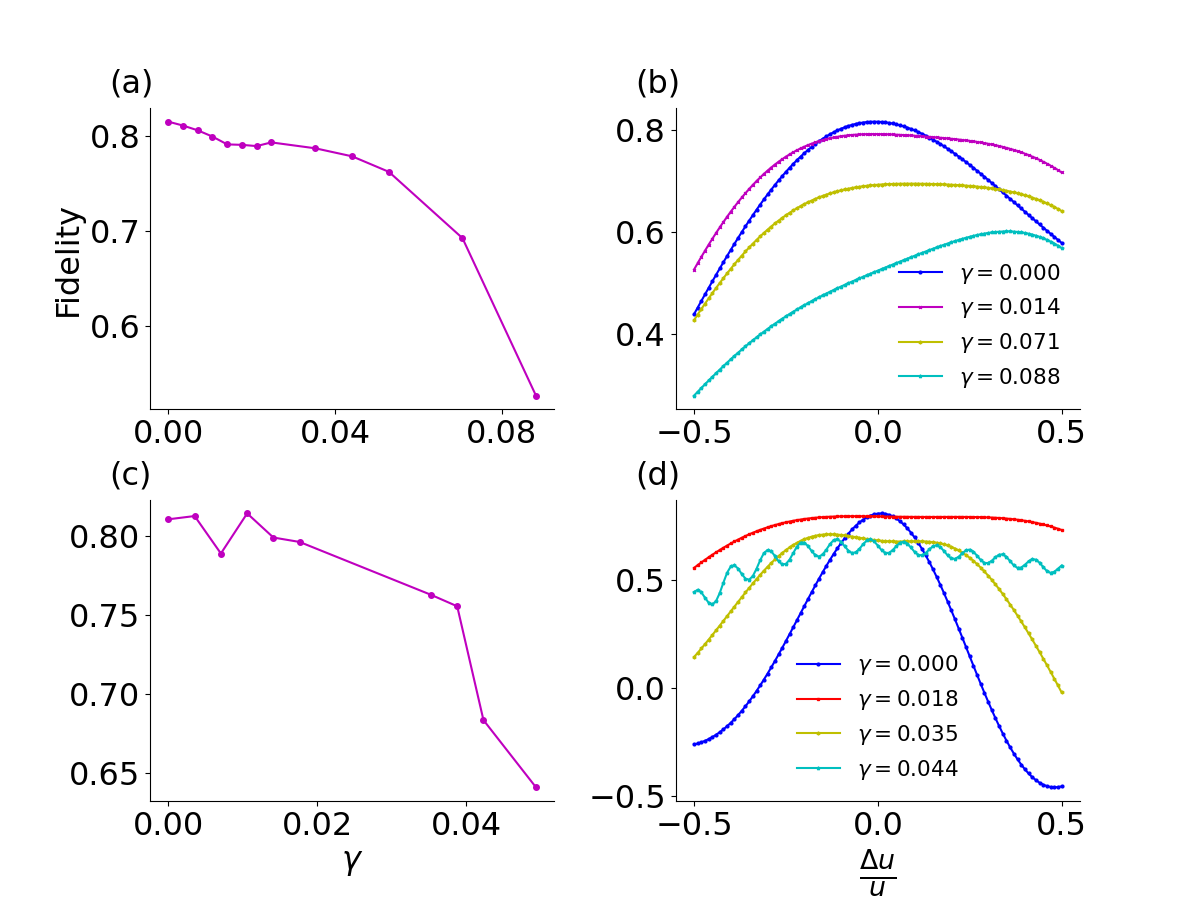}
	\caption{(a,c) LLS preparation fidelity vs noise coefficient $\gamma$ for local (a) and global (c) noise types. (b,d) LLS preparation fidelity vs control field deviation $\Delta u/u$ for local (b) and global (d) noise types. }
	\label{PINN_LLS_comp}
\end{figure}

With the above adaptation, we now construct the PINN sequences using the same Hamiltonian parameters as that of TCP and two noise types: (i) local noise operators 
$V_1 = I_{1x}$, $V_2 = I_{1y}$,$V_3 = I_{2x}$, $V_4 = I_{2y}$, and (ii) global noise operators 
$V_1 = I_{1x} + I_{2x}$, $V_2 = I_{1y} + I_{2y}$.    Fig. \ref{PINN_LLS_comp} (a) and (c) plot the fidelity of LLS vs the noise parameter $\gamma$ for local and global noises, respectively.  We find that the global noise is more effective in lowering the fidelity.  For local noise, we find fairly good fidelity for up to $\gamma=0.07$, while for global noise, the fidelity drops drastically beyond 0.04.  In addition to making the control sequence robust against noise, the above Lindbladian approach makes the control sequence robust against deviations in the control field, either due to miscalibration or field inhomogeneity.  Fig. \ref{PINN_LLS_comp} (b) and (d) plot the fidelity of LLS vs control field deviation $\Delta u/u$ for various values of noise coefficient $\gamma$.  We find significantly improved robustness in both cases with a small value of $\gamma$, while we see adverse effects for larger values of $\gamma$.  In the case of global noise, we also observed an oscillatory profile for larger values of $\gamma$.

\section{Summary and outlook}
\label{sec:summary}
Elaborate, precise, and robust control over quantum dynamics is important  
in all quantum technology applications.  Given the ever-growing capabilities of machine learning, it is a natural pick for the complex challenge of generating quantum controls.  
Generally, a quantum control sequence involves time-dependent Hamiltonians corresponding to various control fields, which makes finding exact propagators challenging.  Most numerical methods for quantum control design rely on 
piece-wise constant approximation renders the output control sequence coarse, higher-bandwidth, and hardware-dependent.  

In this work, we employ a physics-informed neural network (PINN) that overcomes the above problem by encoding the entire shape of the control fields in terms of nonlinear functions characterized by neural network parameters. 
As a result, PINN generates smooth controls, eliminating the need for prior discretization, and therefore, the control sequence can be adapted to any hardware with its optimum timing resolution.

Here, we used PINN to design a quantum gate such as CNOT and prepare a quantum state such as the long-lived singlet (LLS) order in NMR. 
Using two-qubit NMR registers, we experimentally confirmed the good performance of CNOT gate and LLS preparation. We have numerically analyzed the PINN-generated sequences regarding bandwidth and discretization level.  We have also described a Lindbladian-based approach for generating robust quantum control sequences against a general noise source.  This approach also makes the control sequence robust against control field errors such as miscalibration or field inhomogeneities.
We hope this work contributes towards advancing the versatility and effectiveness of quantum control protocols based on machine learning.

\section*{Acknowledgments}
P.B. acknowledges support from the
Prime Minister’s Research Fellowship (PMRF) of the Government of India.
The DST/ICPS/QuST/2019/Q67 funding is acknowledged. We also thank the National Mission on Interdisciplinary Cyber-Physical Systems for funding from the DST, Government of India, through the I-HUB Quantum Technology Foundation, IISER-Pune.
 

\bibliography{references.bib}

\end{document}